\documentclass[12pt]{iopart}
\input{amssym}
\usepackage{amsbsy}
\global\arraycolsep=2pt

\newcommand{\be}{\begin{equation}} 
\newcommand{\ee}{\end{equation}}
\newcommand{\bea}{\begin{eqnarray}}
 \newcommand{\eea}{\end{eqnarray}}

\begin{document}

\title[Green's dyadic approach of the self-stress on a dielectric-diamagnetic cylinder]{Green's dyadic approach of the self-stress on a dielectric-diamagnetic cylinder with non-uniform speed of light}
\vspace{-.2cm}
\author{I Cavero-Pel\'aez  and KA Milton\footnote{On sabbatical leave at the Department of Physics, Washington University, St. Louis, MO 63130 USA}}
\address{Oklahoma Center for High Energy Physics and Homer L. Dodge Department of Physics and Astronomy,University of Oklahoma, Norman, OK 73019 USA}
\eads{\mailto{cavero@nhn.ou.edu}, \mailto{milton@nhn.ou.edu}}


\vspace{-.2cm}
\begin{abstract} 
We present a Green's dyadic formulation to calculate the Casimir energy for a dielectric-diamagnetic
 cylinder with the speed of light differing on the inside and outside.
 Although the result is in general divergent, special cases are meaningful.
It is pointed out how the self-stress on a purely dielectric cylinder vanishes through second order in the deviation of the permittivity from
 its vacuum value, in agreement with the result calculated from the sum
 of van der Waals forces.
\end{abstract}
\vspace{-.5cm}

\vspace{-.1cm}
\section{Formulation of the Green's dyadic approach}
\vspace{-.1cm}
The electromagnetic Green's dyadic functions \cite{schdermil} have been succcesfully used in many occasions (for an extensive view see \cite{miltonbook} and references whitin) and can be applied to very complicated geometries.  Their use  happen to be critical in this calculation \cite{ines&kim}. This approach helps us compute the vacuum expectation value of the fields rigorously; we show that the approach is both illuminating of the physics and unambiguous.
\subsection{Green's dyadic equations; formalism}
In a medium of constant electric permittivity $\varepsilon'$ and magnetic permeability $\mu'$ we insert an infinitely long cylinder of radius $a$ with permittivity and permeability $\varepsilon$ and $\mu$. The product of these parameters is different than that of the outside parameters. There are no real charges of any kind present in the problem, $\rho=\mathbf{J}=0$ and since we work at a fixed frequency we can Fourier transform the electric and magnetic fields,
\be
\mathbf{E}(\mathbf{r},t)= \int_{\infty}^{\infty}\frac{d\omega}{2\pi}\,\mathbf{E}(\mathbf{r},\omega)e^{-i\omega t},\qquad
\mathbf{B}(\mathbf{r},t)= \int_{\infty}^{\infty}\frac{d\omega}{2\pi}\,\mathbf{B}(\mathbf{r},\omega)e^{-i\omega t},
\ee
and the corresponding Maxwell's equations are
\numparts
\begin{eqnarray}
\boldsymbol{\nabla}\times\mathbf{E}= i\omega\mu\mathbf{H},\qquad \boldsymbol{\nabla\cdot}\mathbf{D}=0,\label{elect-maxeq}\\
\boldsymbol{\nabla}\times\mathbf{H}= -i\omega\varepsilon\mathbf{E},\,\quad\,\,\boldsymbol{\nabla\cdot} \mathbf{B}=0.\label{mag-maxeq}
\end{eqnarray}
\endnumparts

In order to write down the Green's dyadic equations, we introduce a 
polarization source $\mathbf{P}$. The first equation in (\ref{mag-maxeq}) and the second one in (\ref{elect-maxeq}) get then changed to,
\be
\boldsymbol{\nabla}\times\mathbf{H}=-i\omega\varepsilon\mathbf{E}-i\omega\mathbf{P},\qquad
\boldsymbol{\nabla\cdot} \mathbf{D}=-\boldsymbol{\nabla\cdot}\mathbf{P}\label{newmag-maxeq}.
\ee
The linear relation of polarization source with the electric field 
defines the Green's dyadic as
\begin{equation}
\mathbf{E}(x)=\int (d x')
\mathbf{\Gamma}(x,x')\boldsymbol{\cdot}\mathbf{P}(x').
\end{equation}
Since the response is translationally invariant in time, we work with the Fourier transform of the dyadic at a given frequency $\omega$.
We can then, by simple substitution, write the  dyadic Maxwell's equations in a medium characterized by 
a dielectric constant $\varepsilon$ and a permeability $\mu$\footnote{In order to have divergenceless Green dyadics, we redefine the electric Green's dyadic in the following way, $\mathbf{\Gamma'(r,r'},\omega)=\mathbf{\Gamma}(\mathbf{r,r'},\omega)
+\frac{\mathbf{1}}{\varepsilon(\omega)}\delta(\mathbf{r-r'})$ and $\boldsymbol\Phi$ is the magnetic dyadic.}:
\numparts
\begin{eqnarray}
\boldsymbol{\nabla}\times\mathbf{\Gamma'}-\mathrm{i}\omega\mu(\omega)\mathbf{\Phi}
&=&\frac{1}{\varepsilon(\omega)}\boldsymbol{\nabla}\times\mathbf{1},\quad\qquad\boldsymbol{\nabla\cdot\Phi}=\mathbf{0},\label{maxeq1}\\
-\boldsymbol{\nabla}\times\mathbf{\Phi}-\mathrm{i}
\omega\varepsilon(\omega)\mathbf{\Gamma'}&=&\mathbf{0},\qquad\qquad\qquad\boldsymbol{\nabla\cdot\Gamma'}=\mathbf{0}.
\label{maxeq2}
\end{eqnarray}
\endnumparts
and where the unit dyadic $\mathbf{1}$ includes a three-dimensional $\delta$ 
function, $\mathbf{1}=\mathbf{1}\delta(\mathbf{r-r'})$.
Quantum mechanically, these Green's dyadics give the one-loop vacuum 
expectation values of the product of fields at a given frequency $\omega$,
\be
\fl\quad\quad\langle\mathbf{E}\mathbf{(r)}\mathbf{E}\mathbf{(r')}\rangle
=\frac{\hbar}{\mathrm{i}}\mathbf{\Gamma(r,r')},
\qquad\langle\mathbf{H}\mathbf{(r)}\mathbf{H}\mathbf{(r')}\rangle
=-\frac{\hbar}{\mathrm{i}}\frac{1}{\omega^2\mu^2}
\overrightarrow{\boldsymbol{\nabla}}
\times\boldsymbol{\Gamma}(\mathbf{r,r'})\times
\overleftarrow{\boldsymbol{\nabla'}}.\label{vev}
\ee
Thus, from the knowledge of the classical Green's dyadics, we can calculate 
the vacuum energy or stress.

Since the TE and TM modes do not separate, we cannot use the general waveguide decomposition of modes into those of TE and TM type\footnote{For example as given in Ref.~\cite{schw&milton2004}.  However, this is here impossible because the TE and TM modes do not separate.  See Ref.~\cite{stratton}.}. However we can introduce the appropriate partial wave decomposition for a cylinder,
in terms of cylindrical coordinates $(r,\theta,z)$\footnote{A slight modification of that given for a conducting cylindrical shell \cite{deraad&milton1981}.}:
\numparts
\begin{eqnarray}
\mathbf{\Gamma'}(\mathbf{r,r'};\omega)&=&\sum_{m=-\infty}^{\infty}
\int_{-\infty}^{\infty}\frac{dk}{2\pi}\bigg{\{}(\boldsymbol{\nabla}\times
\mathbf{\hat z})f_{m}(r;k,\omega)\chi_{mk}(\theta,z)
\nonumber\\
&&\qquad\qquad\mbox{}+\frac{\mathrm{i}}{\omega\varepsilon}\boldsymbol{\nabla}
\times(\boldsymbol{\nabla}\times\mathbf{\hat 
z)}g_m(r;k,\omega)\chi_{mk}(\theta,z)\bigg{\}},\label{gamma}\\
\mathbf{\Phi}(\mathbf{r,r'};\omega)&=&\sum_{m=-\infty}^{\infty}
\int_{-\infty}^{\infty}\frac{dk}{2\pi}\bigg{\{}
(\boldsymbol{\nabla}\times\mathbf{\hat z})\tilde g_{m}(r;k,\omega)
\chi_{mk}(\theta,z)
\nonumber\\
&&\qquad\qquad\mbox{}-\frac{\mathrm{i}\varepsilon}{\omega\mu}\boldsymbol{\nabla}
\times(\boldsymbol{\nabla}\times\mathbf{\hat 
z)}\tilde f_m(r;k,\omega)\chi_{mk}(\theta,z)\bigg{\}},\label{phi}
\end{eqnarray}
\endnumparts
where the cylindrical harmonics are $\chi(\theta,z)=\frac{1}{\sqrt{2\pi}}e^{\mathrm{i}m\theta}e^{\mathrm{i}kz}$, and the dependence of $f_m$ etc.\ on $\mathbf{r'}$ is implicit. Notice that 
these are vectors in the second tensor index. Because of the 
presence of these harmonics we have
\be
\fl\quad\boldsymbol{\nabla}\times\mathbf{\hat z}\rightarrow\mathbf{\hat r}
\frac{\mathrm{i}m}{r}-\boldsymbol{\hat\theta}\frac{\partial}{\partial r}\equiv\boldsymbol{\mathcal{M}},\quad\mbox{and}\quad\boldsymbol{\nabla}\times(\boldsymbol{\nabla}\times\mathbf{\hat z})
\rightarrow\mathbf{\hat 
r}\mathrm{i}k\frac{\partial}{\partial r}-\boldsymbol{\hat \theta}\frac{mk}{r}
-\mathbf{\hat 
z}d_m\equiv\boldsymbol{\mathcal{N}},
\ee
in terms of the cylinder operator $d_m=\frac{1}{r}\frac{\partial}{\partial r}r\frac{\partial}{\partial 
r}-\frac{m^2}{r^2}$.
It is trivial to see that the divergence of (\ref{gamma}) and (\ref{phi}) is zero, satisfying immediately two the the dyadic Maxwell's equations.
Now, if we use the Maxwell equation (\ref{maxeq2}) we conclude\footnote{The 
ambiguity in solving for these equations is absorbed in the definition of 
subsequent constants of integration.}
\be
\tilde g_m=g_m\qquad\mbox{and}\qquad
(d_m-k^2)\tilde f_m=-\omega^2\mu f_m.
\ee

More elaborate work is needed to get a condition form the other Maxwell equation (\ref{maxeq1}). Using the above we can write (\ref{maxeq1}) as,
\begin{eqnarray}
\sum_m\int_{-\infty}^{\infty}\frac{dk}{2\pi}\bigg{\{}-\boldsymbol{\mathcal{M}}\frac{(d_m-k^2)}{\omega^2\mu}\tilde f_m-\frac{\mathrm{i}}{\omega\varepsilon}(d_m-k^2)\boldsymbol{\mathcal{N}}g_m\bigg{\}}&\chi_{mk}(\theta,z)=\nonumber\\
\sum_m\int_{-\infty}^{\infty}\frac{dk}{2\pi}\bigg{\{}\mathrm{i}\omega\mu\boldsymbol{\mathcal{N}}g_m+\varepsilon\boldsymbol{\mathcal{M}}\tilde f_m\bigg{\}}\chi_{mk}(\theta,z)+\frac{1}{\varepsilon}\boldsymbol{\nabla\times1}&.
\end{eqnarray}
if we multiply the above by the expression $\int_0^{2\pi}\int_{-\infty}^{\infty}d\theta dz\chi_{m'k'}^*(\theta,z)$, and apply $\int_0^{2\pi}\int_{-\infty}^{\infty}d\theta dz\chi_{m'k'}^*(\theta,z)\chi_{mk}(\theta,z)=2\pi\delta(k-k')\delta_{mm'}$,
we find
\begin{eqnarray}
-\frac{1}{\omega^2\mu}\boldsymbol{\mathcal{N}}(d_m-k^2+\omega^2\mu\varepsilon)\tilde f_m-\frac{i}{\omega\varepsilon}\boldsymbol{\mathcal{M}}(d_m-k^2+\omega^2\mu\varepsilon)g_m=\nonumber\\
\frac{1}{\varepsilon}\int_0^{2\pi}\int_{-\infty}^{\infty}d\theta dz\chi_{mk}^*(\theta,z)(\boldsymbol{\nabla\times1})\frac{1}{r}\delta(r-r')\delta(\theta-\theta')\delta(z-z'),
\end{eqnarray}
where the delta functions are now made explicit.
By dotting this expression with $\mathbf{\hat{z}}$ we notice that $\mathbf{\hat{z}}\boldsymbol{\cdot\mathcal{M}}=0$ and $\mathbf{\hat{z}}\boldsymbol{\cdot\mathcal{N}}=-d_m$ and after a little manipulation we get to the fourth order differential equation:
\begin{equation}
d_m\mathcal{D}_m\mathbf{\tilde 
f}_m(r;r',\theta',z')=\frac{\omega^2\mu}{\varepsilon}
\boldsymbol{\mathcal{M}'^*}\frac{1}{r}\delta(r-r')\chi_{mk}^*(\theta',z').
\end{equation}
If we now dot it with $(\boldsymbol{\nabla\times}\mathbf{\hat z})$, we learn that a similar equation holds for $g_m$:
\begin{equation}
d_m\mathcal{D}_m\mathbf{
g}_m(r;r',\theta',z')=-\mathrm{i}\omega\boldsymbol{\mathcal{N}'^*}\frac{1}{r}\delta(r-r')\chi_{mk}^*(\theta',z'),
\end{equation}
where we have  made the second, previously suppressed, position arguments explicit and the prime on the differential operator signifies action on the second primed argument\footnote{The Bessel operator appears, $\mathcal{D}_m=d_m+\lambda^2;\quad\lambda^2=\omega^2\varepsilon\mu-k^2$.}.

To solve those equations, we separate variables in the second argument,
\numparts
\begin{eqnarray}
\mathbf{\tilde 
f}_m(r,\mathbf{r'})=\left[\boldsymbol{\mathcal{M}'^*}F_m(r,r';k,\omega)
+\frac{1}{\omega}\boldsymbol{\mathcal{N}'^*}\tilde 
F_m(r,r';k,\omega)\right]\chi_{mk}^*(\theta',z'),
\label{ftilde}\\
\mathbf{g}_m(r,\mathbf{r'})=\left[-\frac{\mathrm{i}}{\omega}
\boldsymbol{\mathcal{N}'^*}G_m(r,r';k,\omega)
-\mathrm{i}\boldsymbol{\mathcal{M}'^*}\tilde 
G_m(r,r';k,\omega)\right]\chi_{mk}^*(\theta',z'),
\label{g}
\end{eqnarray}
\endnumparts
where we have introduced the two scalar Green's functions $F_m, G_m$ satisfying
\be
\fl\qquad d_m\mathcal{D}_mF_m(r,r')=\frac{\omega^2\mu}{\varepsilon}\frac{1}{r}
\delta(r-r'),
\quad\mbox{and}\quad
d_m\mathcal{D}_mG_m(r,r')=\omega^2\frac{1}{r}\delta(r-r'),\label{difeqF&G}
\ee
while  $\tilde F_m$ and $\tilde G_m$ are annihilated by the operator
$d_m\mathcal{D}_m$,
\begin{equation}
d_m\mathcal{D}_m\tilde F(r,r')=d_m\mathcal{D}_m\tilde 
G(r,r')=0.
\label{homodifeq}
\end{equation}

\subsection{Green's dyadic solutions}
The Green's dyadics have now the form:
\numparts
\begin{eqnarray}
\fl\mathbf{\Gamma'(r,r'};\omega)=\sum_{m=-\infty}^{\infty}
\int_{-\infty}^{\infty}\frac{d k}{2\pi}\bigg{\{}\boldsymbol{\mathcal{M 
M'^*}}\left(-\frac{d_m-k^2}{\omega^2\mu}\right)F_m(r,r')+\boldsymbol{\mathcal{N N'^*}}\frac{1}{\omega^2\varepsilon}G_m(r,r')\nonumber\\
\fl\quad\quad+\frac{1}{\omega}\boldsymbol{\mathcal{M 
N'^*}}\left(-\frac{d_m-k^2}{\omega^2\mu}\right)\tilde 
F_m(r,r')+\frac{1}{\omega\varepsilon}\boldsymbol{\mathcal{N M'^*}}\tilde 
G_m(r,r')\bigg{\}}\chi_{mk}(\theta,z)\chi_{mk}^*(\theta',z'),
\label{gammaprimedyadic}\\
\fl\mathbf{\Phi(r,r'};\omega)=\sum_{m=-\infty}^{\infty}
\int_{-\infty}^{\infty}\frac{d k}{2\pi}\bigg{\{}
-\frac{\mathrm{i}}{\omega}\boldsymbol{\mathcal{M 
N'^*}}G_m(r,r')-\frac{\mathrm{i}\varepsilon}{\omega\mu}\boldsymbol{\mathcal{N 
M'^*}}F_m(r,r')\nonumber\\
\fl\quad\quad-\mathrm{i}\boldsymbol{\mathcal{M M'^*}}\tilde G_m(r,r')-\frac{\mathrm{i}\varepsilon}{\omega^2\mu}\boldsymbol{\mathcal{N 
N'^*}}\tilde F_m(r,r')\bigg{\}}\chi_{mk}(\theta,z)\chi_{mk}^*(\theta',z').
\label{phidyadic}
\end{eqnarray}
\endnumparts
In the following, we will apply these equations to a dielectric-diamagnetic 
cylinder of radius $a$, where the interior of the cylinder is characterized by 
a permittivity $\varepsilon$ and permeability $\mu$, while the outside is 
vacuum, so $\varepsilon=\mu=1$ there. 
Let us consider the case that the source point is outside, $r'>a$. If the 
field point is also outside, $r,r'>a$, the scalar Green's functions $F'_m, 
G'_m, \tilde F', \tilde G'$  that make up the above Green's dyadics (we 
designate with primes the outside scalar Green's functions or 
constants) obey the differential equations (\ref{difeqF&G}) and 
(\ref{homodifeq}) with $\varepsilon=\mu=1$.
The solutions to these equations are\footnote{For details see \cite{ines&kim} and \cite{mythesis}.}:
\begin{eqnarray}
\fl\qquad F'_m(r,r')&=&\frac{\omega^2}{\lambda'^2}\left[\frac{a'^F_m}
{r'^{|m|}}+b'^F_mH_m(\lambda' 
r')\right]{r^{-|m|}}-\frac{\omega^2}{\lambda'^2}\frac{1}{2|m|}
\left(\frac{r_<}{r_>}\right)^{|m|}\nonumber\\
\fl &+&\left[\frac{A'^F_m}{r'^{|m|}}+B'^F_mH_m
(\lambda' r')\right]H_m(\lambda' 
r)-\frac{\omega^2}{\lambda'^2}\frac{\pi}{2\mathrm{i}}J_m(\lambda' r_<)
H_m(\lambda' r_>),
\end{eqnarray}
while $G'_m$ has the same form with the constants $a'^F_m, b'^F_m, A'^F_m, 
B'^F_m$ replaced by $a'^G_m,b'^G_m,A'^G_m,B'^G_m$, respectively. The 
homogeneous differential equations have solutions
\begin{equation}
\fl\qquad\tilde F'_m(r,r')=\frac{\omega^2}{\lambda'^2}\left[\frac{a'^{\tilde 
F}_m}{r'^{|m|}}+b'^{\tilde F}_mH_m(\lambda' 
r')\right]{r^{-|m|}}+\left[\frac{A'^{\tilde F}_m}{r'^{|m|}}+B'^{\tilde 
F}_mH_m(\lambda' r')\right]H_m(\lambda' r),
\label{tildeF}
\end{equation}
while in  $\tilde G'_m$ we replace 
$a'^{\tilde F}\rightarrow a'^{\tilde G}$, etc. 

When the source point is outside and the field point is inside, all the
Green's functions satisfy the
homogeneous equations (\ref{homodifeq}) with $\varepsilon$, $\mu\ne1$, 
and then $F_m, G_m, \tilde F_m, \tilde G_m,$ are of the same form as in equation (\ref{tildeF}) with the corresponding change 
of constants. In all of the above, the outside and inside forms of 
$\lambda$ are given by $\lambda'^2=\omega^2-k^2$ and
$\lambda^2=\omega^2\mu\varepsilon-k^2$.

The various constants are to be determined, as far as possible, by the 
boundary 
conditions at $r=a$. The boundary conditions at the surface of the dielectric 
cylinder are the continuity of tangential components of the electric field, of 
the normal component of the electric displacement, of the normal component of 
the magnetic induction, and of the tangential components of the magnetic field
(we assume that there are no surface charges or currents). In terms of the Green's dyadics, the conditions read
\numparts
\begin{eqnarray}
\mathbf{\hat r}\boldsymbol{\cdot}\varepsilon\boldsymbol{\Gamma'}
\bigg |_{r=a-}^{r=a+}&=&\mathbf{0},\qquad
\boldsymbol{\hat\theta}\boldsymbol{\cdot\Gamma'}\bigg |_{r=a-}^{r=a+}=\mathbf{0},\qquad
\mathbf{\hat z}\boldsymbol{ \cdot\Gamma'}\bigg |_{r=a-}^{r=a+}
=\mathbf{0},\label{bc3}\\
\mathbf{\hat r}\boldsymbol{\cdot}\mu\boldsymbol{\Phi}\bigg|_{r=a-}^{r=a+}&=&
\mathbf{0},\qquad
\boldsymbol{\hat\theta}\boldsymbol{\cdot\Phi}\bigg |_{r=a-}^{r=a+}=
\mathbf{0},\qquad
\mathbf{\hat z}\boldsymbol{\cdot\Phi}\bigg |_{r=a-}^{r=a+}=\mathbf{0}.\label{bc6}
\end{eqnarray}
\endnumparts
By imposing those boundary conditions, we find that the only constants contributing to the energy are:
\numparts
\begin{eqnarray}
B^{\tilde 
G}_m&=&-\frac{\varepsilon^2}{\mu}(1-\varepsilon\mu)
\frac{mk\omega}{\lambda\lambda'D}J_m(\lambda 
a)H_m(\lambda'a)B^F_m,\\
B'^{\tilde 
G}_m&=&-\left(\frac{\lambda}{\lambda'}\right)^2
\frac{\varepsilon}{\mu}(1-\varepsilon\mu)
\frac{mk\omega}{\lambda\lambda'D}J^2_m(\lambda 
a)B^F_m,\\
B'^F_m&=&\frac{\omega^2}{\lambda'^2}\frac{\pi}{2\mathrm{i}}\frac{J_m(\lambda'a 
)}{H_m(\lambda'a)}+\left(\frac{\lambda}{\lambda'}\right)^2
\frac{\varepsilon}{\mu}\frac{J_m(\lambda 
a)}{H_m(\lambda'a)}B_m^F,\\
B^{\tilde 
F}_m&=&-\frac{\mu}{\varepsilon^2}(1-\varepsilon\mu)
\frac{mk\omega}{\lambda\lambda'\tilde 
D}J_m(\lambda a)H_m(\lambda'a)B^G_m,\\
B'^{\tilde 
F}_m&=&-\left(\frac{\lambda}{\lambda'}\right)^2
\frac{1}{\varepsilon}(1-\varepsilon\mu)\frac{mk\omega}{\lambda\lambda'\tilde 
D}J^2_m(\lambda a)B^G_m\label{bprimetilf},\\
B'^G_m&=&\frac{\omega^2}{\lambda'^2}\frac{\pi}{2\mathrm{i}}\frac{J_m(\lambda'a 
)}{H_m(\lambda'a)}+\left(\frac{\lambda}{\lambda'}\right)^2
\frac{1}{\varepsilon}\frac{J_m(\lambda 
a)}{H_m(\lambda'a)}B_m^G,
\end{eqnarray}
\endnumparts
all in terms of  $B_m^F=-\frac{\mu}{\varepsilon}\frac{\omega^2}{\lambda\lambda'} \frac{D}{\Xi} \quad\mbox{and}\quad B_m^G=-\varepsilon\frac{\omega^2}{\lambda\lambda'}\frac{\tilde D}{\Xi}$.

The denominators occurring here are\footnote{The denominator structure appearing in $\Xi$ is precisely that given by Stratton \cite{stratton},and is the basis for the calculation given, for example in Ref.~\cite{milt-nest-nest}.  It is also employed in an independent rederivation of the Casimir energy for a dilute dielectric cylinder \cite{romeom}.}
\numparts
\begin{eqnarray}
\Xi&=&(1-\varepsilon\mu)^2\frac{m^2k^2\omega^2}{\lambda^2\lambda^{\prime2}}
J^2_m(\lambda  a)H^2_m(\lambda'a)-D\tilde D,\\
D&=&\varepsilon\lambda'aJ'_m(\lambda a)H_m(\lambda'a)-\lambda 
aH'_m(\lambda'a)J_m(\lambda a),\\
\tilde D&=&\mu\lambda'aJ'_m(\lambda a)H_m(\lambda'a)-\lambda 
aH'_m(\lambda'a)J_m(\lambda a).
\end{eqnarray}
\endnumparts

It is now easy to check that the terms in the Green's functions that involve powers of $r$ 
or $r'$ do not contribute to the electric or magnetic fields.
 So, even though we are not able to determine all the constants (notice that there is some ambiguity in these since they cannot be uniquely determined), it is not an issue since the energy will be
 well defined \cite{ines&kim, mythesis}. These constants enter always in the same form and 
therefore their individual values are not relevant. As we might have anticipated, only the pure Bessel function terms contribute. It might be thought that $m=0$ is a special case, and indeed $\frac{1}{2|m|}\left(\frac{r_<}{r_>}\right)^{|m|}\rightarrow\frac{1}{2}\ln\frac{r_<}{r_>}$, but just as the latter is correctly interpreted as the limit as $|m|\rightarrow0$, so the coefficients in the Green's functions turn out to be just the $m=0$ limits for those given above, so the $m=0$ case is properly incorporated.

\subsection{Stress on the cylinder}
We are now in a position to calculate the pressure on the surface of the 
cylinder from the radial-radial component of the stress tensor
\begin{equation}
P=\langle T_{rr}\rangle(a-)-\langle T_{rr}\rangle(a+)
\end{equation}
where $T_{rr}=\frac{1}{2}\left[\varepsilon(E^2_{\theta}+E^2_z-E^2_r)
+\mu(H^2_{\theta}+H^2_z-H^2_r)\right]$.
As a result of the boundary conditions, the pressure on the 
cylindrical walls is given by the expectation value of the squares of field 
components just outside the cylinder, therefore
\be
\fl\qquad T_{rr}\big|_{a-}-T_{rr}\big|_{a+}=\frac{\varepsilon-1}{2}
\left(E^2_{\theta}+E^2_z+\frac{E^2_r}{\varepsilon}\right)\bigg|_{a+}
+\frac{\mu-1}{2}\left(H^2_{\theta}+H^2_z
+\frac{H^2_r}{\mu}\right)\bigg|_{a+},\label{pressure}
\ee
where the expectation values are given by (\ref{vev}) in terms of the Green's functions. We obtain the pressure on the cylinder as
\begin{eqnarray}
\fl P=&&\frac{\varepsilon-1}{16\pi^3a^4}\sum_{m=-\infty}^{\infty}
\int_{-\infty}^{\infty}d\zeta 
a\,dka\frac{\hbar}{\tilde\Xi}\bigg{\{}K'^2_m(y')I_m(y)I'_m(y)y(k^2a^2
-\zeta^2a^2\mu)-K'_m(y')I^2_m(y)\nonumber\\
\fl&\times&K_m(y')\bigg[\frac{m^2k^2a^2\zeta^2a^2}
{y'^3\varepsilon}\bigg(-2(\varepsilon+1)(1-\varepsilon\mu)+\frac{k^2a^2-\zeta^2a^2\varepsilon}{y^2}
(1-\varepsilon\mu)^2\bigg)\nonumber\\
\fl&-&\frac{y^2}{y'}\left(\frac{m^2}{y'^2}\left(k^2a^2
-\frac{\zeta^2a^2}{\varepsilon}\right)+y'^2\right)\bigg]-K'_m(y')I'^2_m(y)K_m(y')\mu y'(k^2a^2-\zeta^2a^2\varepsilon)\nonumber\\
\fl&-&I_m(y)I'_m(y)K^2_m(y')y\left[\frac{m^2}{y'^2}(k^2a^2\mu
-\zeta^2a^2)+y'^2\mu\right]\bigg{\}}+\{(\varepsilon\leftrightarrow\mu)\},\label{rotpressure}
\end{eqnarray} 
where we have performed the Euclidean rotation $\omega\rightarrow \mathrm{i}\zeta ,\lambda\rightarrow \mathrm{i}\kappa$, and $\tilde\Xi$ is the rotated $\Xi$. Here $y=\kappa a$, $y'=\kappa'a$ and the last bracket indicates that the 
expression there is similar to the one for the electric part by switching 
$\varepsilon$ and $\mu$, showing manifest symmetry between the electric and 
magnetic parts. However, this expression is incomplete. It contains an 
unobservable ``bulk'' energy contribution, which the formalism would give if 
either medium, that of the interior with dielectric constant $\varepsilon$ and 
permeability $\mu$, or that of the exterior with dielectric constant and 
permeability unity, fills all the space \cite{miltonng97}. The corresponding stresses are 
computed from the free Green's functions which satisfy (\ref{difeqF&G}), and have solutions
\begin{equation}
\fl\qquad F^{(0)}_m(r,r')=\frac{\mu}{\varepsilon}G^{(0)}_m(r,r')=
-\frac{\omega^2\mu}{\varepsilon\lambda^2}\left[\frac{1}{2|m|}
\left(\frac{r_<}{r_>}\right)^{|m|}+\frac{\pi}{2\mathrm{i}}J_m(\lambda 
r_<)H_m(\lambda r_>)\right],
\label{freegreensfunct}
\end{equation}
where $0<r,r'<\infty$. Notice that in this case, both $\tilde F^{(0)}_m$ and 
$\tilde G^{(0)}_m$ are zero. After the Euclidean rotation the bulk pressure becomes 
\begin{eqnarray}
\fl P^b=&T^{(0)}_{rr}&(a-)-T^{(0)}_{rr}(a+)=\frac{\hbar}{16\pi^3a^4}\sum_{m=-\infty}^{\infty}\int_{-\infty}^{\infty}
d\zeta a\, d k a\big{\{}y^2I'_m(y)K'_m(y)\nonumber\\
\fl&-&(y^2+m^2)I_m(y)K_m(y)-y'^2I'_m(y')K'_m(y')+(y'^2+m^2)I_m(y')K_m(y')
\big{\}}.
\label{bpressure}
\end{eqnarray}
This term must be subtracted from the pressure given in (\ref{rotpressure}). 
Note that $P^b=0$ in the special case $\varepsilon\mu=1$ as it should be.
\section{Dilute dielectric cylinder}
We now turn to the case of a dilute dielectric medium filling the cylinder, 
that is, set $\mu=1$ and consider $\varepsilon-1$ as small. We can then expand 
the integrand in (\ref{rotpressure}) and (\ref{bpressure}) in powers of $(\varepsilon-1)$. Because the expression in (\ref{rotpressure}) is already proportional to that factor, we need only 
expand the integrand to first order. The {\emph{total}} pressure can then be written as:
\begin{eqnarray}
\fl P-P^b&=&\frac{\hbar}{8\pi^2a^4}(\varepsilon-1)^2\sum_{m=-\infty}^{\infty}
\int_{0}^{\infty}d y\bigg{\{}\frac{y^4}{2}\bigg[\frac{1}{2}
K'^2_m(y)I'_m(y)I_m(y)\nonumber\\
\fl&+&K'^2_m(y)I'^2_m(y)\frac{y}{4}-K'^2_m(y)I^2_m(y)\frac{y}{4}\left(1+\frac{m^2}{y^2}\right)+K'_m(y)I'^2_m(y)K_m(y)\nonumber\\
\fl&+&K^2_m(y)I^2_m(y)\frac{y}{2}\left(1+\frac{m^2}{y^2}\right)
\left(1-\frac{m^2}{2y^2}\right)-K^2_m(y)I'^2_m(y)\frac{y}{2}\left(1-\frac{m^2}{2y^2}\right)\nonumber\\
\fl&+&K^2_m(y)I'_m(y)I_m(y)\left(1+\frac{m^2}{2y^2}\right)
\bigg]+\frac{3y}{16}[I_m(y)K_m(y)]'\bigg{\}}.
\label{p1st&2ndorder}
\end{eqnarray}
 Thus the total stress vanishes in leading order which is consistent with the interpretation of the Casimir energy as arising 
from the pairwise interaction of dilutely distributed molecules.
Several methods to compute this integral are explain with great detail in \cite{ines&kim} and in \cite{mythesis}. There it is shown that making use of the asymptotic expansion for the Bessel functions, we can numerically evaluate the integral 
\begin{eqnarray}
\fl P&=&\frac{(\varepsilon-1)^2}{32\pi^2a^4}(
-0.007612+0.287168+0.024417-0.002371-0.000012-0.301590)\nonumber\\
\fl&=&0.000000,
\label{zresult}
\end{eqnarray}
and by introducing an exponential regulator $e^{-\delta y}$ in (\ref{p1st&2ndorder}) we can unambiguously separate the two divergent terms
\begin{equation}
P_{\rm div}=\frac{(\varepsilon-1)^2}{32\pi^2a^4}
\left(\frac{13\pi^2}{32\delta^3}-\frac{315\pi}{8192\delta}\right).
\end{equation}
The form of the divergences is exactly as expected \cite{bordag01,barton}.
In particular, there is no $1/\delta^2$ divergence.
 How do we interpret these
terms?  It is perhaps easiest to imagine that $\delta$ as given in terms of
a proper-time cutoff, $\delta=\tau/a$, $\tau\to 0+$.  Then if we consider the
energy, rather than the pressure, the divergent terms have the form $E_{\rm div}=e_3\frac{aL}{\tau^3}+e_1 \frac{L}{a}\frac{1}{\tau}$.
Here $L$ is the (large) length of the cylinder. 
 Thus, the leading divergence
corresponds to an energy term proportional to the surface of the cylinder,
and it therefore appears sensible to absorb it into a renormalized surface
energy which enters into a phenomenological description of the material system.
The $1/\tau$ divergence is more problematic.  It is proportional to the
ratio of the length to the diameter of the cylinder, so it seems likely that
this would be interpretable as an energy term referring to the shape of the
body.
In any case, although the structure of the divergences is universal,
the coefficients of those divergences depend in detail upon the particular
regularization scheme adopted. The nature of divergences in such Casimir calculations is still under active study \cite{miltonbook,jaffe,fulling,in-mil-wag}. In contrast, the term proportional to $(\varepsilon-1)^2/a^2$ is unique. The universality of the
finite Casimir term makes it hard not to think it has some real significance.
 Thus, of course, it could not have been
any other than that zero value given by the van der Waals calculations 
\cite{milt-nest-nest, romeo, milonni}. 
\vspace{-.2cm} 
\section{Conclusion}
We have shown how the Green's dyadic formulation, modified for dielectric materials, exhibits a transparent way to calculate the Casimir energies of a dielectric-diamagnetic cylinder and showed that in the dilute case, it coincides with that obtain by summing the van der Waals energies of the constituent molecules. However, the identity is not really that trivial, because both the van der Waals and the Casimir energies contain divergent contributions. This is particularly crucial when one is considering the self-stress of a single body rather than the energy of interaction of distinct bodies. It was nontrivial to show the analog for the case of the dielectric sphere \cite{ddsph}, and the calculation for the dielectric cylinder turned out to be extraordinarily difficult. 
\vspace{-.4cm}
\section*{Acknowledgment}
\vspace{-.4cm}
We thank the US Department of Energy for partial support of this
research.  We acknowledge numerous communications with August Romeo,
and many helpful conversations with K. V. Shajesh. We are grateful to Emilio Elizalde for making everybody welcome and for his excellent organization of the QFEXT05 workshop.

\vspace{-.4cm}


\end{document}